\documentclass[reprint,twocolumn,aps,prb,superscriptaddress]{revtex4-2}
\usepackage[T1]{fontenc}
\usepackage[utf8]{inputenc}
\usepackage{lmodern}
\usepackage{graphicx}
\usepackage{amsmath}
\usepackage{xcolor}
\usepackage{float}
\usepackage{hyperref}
\usepackage{braket}
\usepackage{amssymb,amsmath,amsthm, array}
\usepackage{mdframed}
\usepackage{systeme,mathtools}

\def  \bsigma   {\mbox{\boldmath$\sigma $}}

\DeclareUnicodeCharacter{0301}{-}

\begin{document}
\renewcommand{\vec}[1]{\mathbf{#1}}
\newcommand{\ii}{\mathrm{i}}
\def\ya#1{{\color{orange}{#1}}}

\title{The controlled rotation of entanglement in altermagnets}

\author{M. Kulig}
\affiliation{Department of Physics and Medical Engineering, Rzesz\'ow University of Technology, 35-959 Rzesz\'ow, Poland}

\author{T. Masłowski}
\affiliation{Department of Physics and Medical Engineering, Rzesz\'ow University of Technology, 35-959 Rzesz\'ow, Poland}

\author{K. A. Kouzakov}
\affiliation{Faculty of Exact and Natural Sciences, Tbilisi State University, Chavchavadze av.3, 0128 Tbilisi, Georgia}

\author{V. K. Dugaev}
\affiliation{Department of Physics and Medical Engineering, Rzesz\'ow University of Technology, 35-959 Rzesz\'ow, Poland}

\author{P. Kurashvili}
\affiliation{National Centre for Nuclear Research, Warsaw 00-681, Poland}

\author{S. Wolski}
\affiliation{Department of Physics and Medical Engineering, Rzesz\'ow University of Technology, 35-959 Rzesz\'ow, Poland}

\author{M. Inglot}
\affiliation{Department of Physics and Medical Engineering, Rzesz\'ow University of Technology, 35-959 Rzesz\'ow, Poland}

\author{C. Jasiukiewicz}
\affiliation{Department of Physics and Medical Engineering, Rzesz\'ow University of Technology, 35-959 Rzesz\'ow, Poland}

\author{L. Chotorlishvili}
\affiliation{Department of Physics and Medical Engineering, Rzesz\'ow University of Technology, 35-959 Rzesz\'ow, Poland}
\affiliation{Faculty of Exact and Natural Sciences, Tbilisi State University, Chavchavadze av.3, 0128 Tbilisi, Georgia}

\date{\today}
\begin{abstract}
Altermagnetism became very popular because of unique features, namely coupling between magnetic properties and momentum of itinerant electrons. The particular model of the altermagnetic system of our interest has already been studied in recent publications in a different context: Phys. Rev. B \textbf{108}, L140408 (2023). Here, we study the scattering process of an itinerant electron from the altermagnetic system on the electron localized in a quantum dot. We found a spatially inhomogeneous distribution of quantum entanglement in the post-scattering state. An interesting observation is the controlled rotation of entanglement achieved by means of spin-orbital coupling constant in altermagnetic. We also studied Reny entropy and the effect of disorder in the system leading to randomness in the spin-orbit constant. Our main finding is that due to the unique properties of an altermagnetic system, tuning the applied external magnetic field allows tailoring of the desired entangled state. Thus, the scattering process, in essence, mimics the Hadamard-CNOT Gate transformation, converting the initial disentangled state into the entangled state of Bell's state. In particular, we achieved more than 70 percent fidelity between the post-scattering and Bell's states.
\end{abstract}

\maketitle

\section{Introduction}

The linearity of the Schr\"odinger equation leads to the superposition principle, a fundamental principle of quantum mechanics. Due to the superposition principle, the linear combinations of solutions also satisfy the Schr\"odinger equation. The superposition principle and quantumness of the system form the basis of the functionality of quantum computers, and it has certain advantages. The list of proposed quantum registers for quantum computers is pretty long. It includes e.g., superconducting quantum computing with Josephson junction qubits, trapped ions, and quantum dots  
 \cite{arakawa2020progress, PhysRevLett.121.110503, PhysRevLett.121.033902, PhysRevB.101.081408, PhysRevB.101.045306,kojima2021probabilistic}. The Quantum register is a repository of qubits with quick access to them and the possibility of controlling their states through the quantum gates. In particular, semiconductor quantum dots attract the attention of the quantum information community \cite{PhysRevB.86.085423, PhysRevB.100.174413, PhysRevB.98.104407}.   Quantum dots (QD) in the materials with a strong spin-orbit (SO)  interaction (graphene,  carbon nanotubes, and topological insulators \cite{PhysRevB.89.075103}) are in the scope of interest \cite{trauzettel2007spin,ezawa2012topological}. Spin-orbit interaction (SOI) can be exploited to increase the fidelity of quantum gates \cite{PhysRevLett.83.4204, PhysRevA.57.120}. Recently, the study of altermagnetism became a hot topic \cite{PhysRevX.12.040501, PhysRevX.12.040002, PhysRevLett.132.036702, PhysRevB.106.094432,PhysRevLett.131.256703}. Altermagnetic materials, such as insulators $\text{FeF}_2$ and $\text{MnF}_2$, semiconductors $\text{MnTe}$, and metals $\text{RuO}_2$, are characterized by magnetic properties strongly coupled with the momentum of the electrons. Coupling between electronic and magnetic properties opens a new possibility for controlling qubits in QDs through the SOI and itinerant electron states. Altermagnets are viewed as a new type of magnetic materials \cite{PhysRevX.12.031042,PhysRevX.12.040501,vsmejkal2022anomalous,PhysRevX.12.040002}. The main peculiarity of the altermagnetic state is the existence of spin splitting of electron energy bands, which is generated by the usual antiferromagnetic ordering in the magnetic sublattice (see, e.g., review articles \cite{krempasky2024altermagnetic,zhu2024observation}). The spin splitting is quite unusual for conventional antiferromagnets, and is due to the specific orientational symmetry of electron orbitals, so that the band electrons moving in different directions experience the magnetization of different magnetic sublattices \cite{PhysRevX.12.040501,PhysRevB.107.L100418,zhu2024observation}. It was found that the symmetry of certain crystals like, e.g., RuO$_2$, MnTe etc. allows such an altermagnetic state \cite{mazin2021prediction,PhysRevB.107.L100418,PhysRevLett.130.036702,PhysRevB.106.195149,feng2022anomalous,PhysRevLett.132.036702}.
	
The spin splitting in altermagnets has some similarity to the spin splitting due to relativistic spin-orbit interaction and, correspondingly, makes it possible to use altermagnets in various spintronic applications. For instance, it was demonstrated the existence of anomalous Hall effect, spin current generation, spin Seebeck and spin Nernst effects, et al. in the altermagnets \cite{PhysRevLett.130.036702,PhysRevB.106.195149,feng2022anomalous,bose2022tilted,
bai2022observation,PhysRevLett.129.137201,PhysRevB.108.L180401}. In this work, we demonstrate that the nonrelativistic coupling of spin to the orbital motion in altermagnets can be also used for manipulating the quantum entanglement by using the scattering of electrons from the quantum dot.

A disentangled bipartite system $\ket{\psi}_{AB}=\ket{0}_A\otimes\ket{0}_B$ can be entangled by performing a unitary quantum gate operation: $\text{Hadamard}_A\circ\text{CNOT}_B\ket{0}_A\ket{0}_B=\ket{\Phi}_{AB}^+$, where $\ket{\Phi}_{AB}^+$ is the Bell's state. 
While the formal procedure is relatively straightforward, its implementation for the realistic physical system is a demanding problem. One practical realization of the entangling gate procedure is the elastic spin-dependent scattering protocol \cite{PhysRevLett.130.221801,afik2022quantum,afik2024entanglement,PhysRevA.100.022311}.
A unitary scattering $\hat S$ matrix converting disentangled state into entangled post-scattering state can be viewed as equivalent of the Hadamard and CNOT gates $\hat S\equiv\left\lbrace \text{Hadamard}_A\circ\text{CNOT}_B\right\rbrace $. It connects the initial disentangled bipartite state with a final entangled state $\ket{\Psi_{AB}^f(\infty)}=\hat S \ket{\Psi_{AB}^i(-\infty)}$.  
In the present work, we study the scattering process of the itinerant electron of the altermagnetic system on the electron localized in the quantum dot, which is deposited on the surface of altermagnet. We prove that the scattering of two electrons leads to the formation of the entangled state due to the features of the altermagnetic system. The key role belongs to the coupling between electronic and magnetic properties and presence of a SOI effect in the altermagnetic system.\vspace{0.2cm}\\
Measuring the entanglement in the experiments is related to specific dilemmas. Let  $\mathcal{\hat O}$ be an arbitrary hermitian quantum operator, and $Tr(\hat \rho\mathcal{\hat O})$ the expectation value of it averaged through the density matrix of the system $\hat\rho$. Then 
nonlinear functions of the density matrix i. e., purity $Tr(\hat \rho^2)$ can be measured directly without reconstruction of the whole density matrix 
$\hat \rho$, \cite{PhysRevA.65.062320,PhysRevLett.89.127902,PhysRevLett.90.167901,PhysRevLett.98.140505}. There are two different direct and indirect measurement schemes. 
The direct measurement scheme implies $n$ identical copies of the quantum systems prepared in the same state $\hat\rho$. Measurements are performed on those multiple copies. In many cases, it is a "thought experiment" since the non-cloning theorem does not permit copying and preparing desired number of the identical quantum states. The elegant solution is a random measurement that only implies a single copy of $\hat\rho$, \cite{PhysRevLett.108.110503, PhysRevLett.120.050406}. One applies the random unitary rotation operator $\hat U$  to a single copy of the state  $\hat U\hat \rho_{AB}\hat U^{-1}$, where $\hat\rho_{AB}=\ket{\Psi}\bra{\Psi}_{AB}$ is the density matrix of the bipartite system. Then, the result is averaged over random unitaries. In essence, the ensemble average of a single random measurement is equivalent to the set of deterministic measurements \cite{PhysRevLett.108.110503, PhysRevLett.120.050406}. In our case, the practical implementation of the random unitary rotation operator is equivalent to the random SOI in the scattering problem. Randomness in the SOI is related to the disorder and impurities and is a realistic assumption for two-dimensional solid-state materials
\cite{PhysRevB.67.161303}. In what follows, we will be focused on Renyi entropies, which can be determined from a tomographic reconstruction of the quantum state \cite{haffner2005scalable, PhysRevLett.105.150401,lanyon2017efficient,torlai2018neural} and measured experimentally \cite{islam2015measuring,kaufman2016quantum}:
$S^{(n)}(\hat\rho_A)=[1/(n-1)]\log Tr(\hat\rho_A)$, where $\hat\rho_A$ is the reduced matrix $\hat\rho_A=Tr_B(\hat\rho_{AB})$ of the part $A$. 
We are interested in outcome probabilities
$P(s)=Tr[\hat U\hat\rho_A\hat U^\dagger\mathcal{P}_s]$, where $\mathcal{P}_s=\ket{s}\bra{s}$ are projectors and $\hat U$ is the random unitary matrix. Then Renyi entropy is extracted through the statistical moments
\begin{eqnarray}\label{statistical moments}
\langle P(s)^n\rangle=\langle Tr[(\hat U\hat\rho_A\hat U^\dagger)^{\otimes n}\mathcal{P}_s^{\otimes n}]\rangle,
\end{eqnarray} 
where $\langle ...\rangle$ means averaging over disorder. 
We consider a bipartite system consisting of an altermagnet (itinerant electron \textbf{A}) and quantum dot (localized electron \textbf{B}) prepared initially in a product state. We assume the randomness of the SO coupling constant (due to the disorder and impurities that may exist in the realistic material). We tackle the elastic scattering process as a random unitary operation and explore the entanglement of the scattered state. Thus random unitaries in our proposal are replaced by a random SOI elastic scattering process $\hat U=\hat S$. We calculate fidelity $\mathcal{F}\left(\ket{\Psi_{AB}^f(\infty)}, \ket{\Psi}_{AB}^G\right)$ between scattered state
$\ket{\Psi_{AB}^f(\infty)}=\hat S \ket{\Psi_{AB}^i(-\infty)}$ and the state obtained through the applied gate $\ket{\Psi}_{AB}^G=\left\lbrace \text{Hadamard}_A\circ\text{CNOT}_B\right\rbrace \ket{\Psi_{AB}^i(-\infty)}$.
The paper is organized as follows: In section \textbf{II}, we solve the scattering problem; in  \textbf{III}, we present results for concurrence obtained for the deterministic SO coupling constant. The Rényi entropy is discussed in section \textbf{IV}. Section \textbf{V} concludes the work. 

\section{Scattering problem}
\label{sec:Scattering}
\subsection{Formulation of the problem}
The total Hamiltonian of the system comprises the three terms
\begin{eqnarray}
\label{The total Hamiltonian of the system}
&& \hat H_{tot}=\hat H_{AM}+\hat H_{D}+\hat V.
\end{eqnarray}
Here $\hat H_{AM}=\varepsilon_k+\alpha_e\hat\sigma_A^zk_xk_y$, $k_x=k\cos\theta$, $k_y=k\sin\theta$ is the low-energy Hamiltonian for the altermagnet, $\varepsilon_k=\hbar^2k^2/2m$, $\hat\sigma_A^z$ is the Pauli operator for the spin of itinerant electron, $k$ is the momentum of itinerant electron, and $\alpha_e$ is the SOI constant \cite{PhysRevB.108.L140408}. The Hamiltonian of electron localized in the quantum dot (QD) has the form:
\begin{eqnarray}\label{Hamiltonian of the electron localized}
&&\hat H_{D}=-B\hat\sigma_B^z-\frac{\hbar^2}{2m}\vec\nabla^2+\frac{1}{2}m\omega_0^2r^2,
\end{eqnarray}
where $\omega_0$ is the frequency of electron oscillation in QD.  
The external magnetic field applied locally to the quantum dot allows it to freeze (strong field) or relax (weak field)
the spin of the localized electron $\hat\sigma_B^z$ depending on the value of Zeeman energy $B\equiv \hbar\gamma_eB$.
The lowest eigenstate of the localized electron has the form 
\begin{eqnarray}\label{confined in the quantum dot}
&&\psi_{D}(\textbf{r})=\frac{1}{l_B\sqrt{\pi}}\, \exp\left(-\frac{x^2+y^2}{2l_B^2}\right) ,
\nonumber\\
&& l_B^2=l_0^2/\sqrt{1+B^2e^2l_0^4/4\hbar^2}.
\end{eqnarray}
Here $l_0=(\hbar/m_e\omega_0)^{1/2}$ is the confinement length \cite{PhysRevB.82.045311}. Hereafter we set dimensionless $\textbf{r}\equiv\textbf{r}/l_0$ and measure the distance in terms of the confinement length.
The last term in Eq.(\ref{The total Hamiltonian of the system}) describes the interaction between localized and itinerant electrons. We consider the short range interaction case \cite{PhysRevB.89.075426}:
\begin{eqnarray}\label{interaction between localized and itinerant electrons}
&&\hat V =J\, \hat\bsigma _A\cdot \hat\bsigma _B\; \delta\left(\textbf{r}_1-\textbf{r}_2\right).
\end{eqnarray}
The coupling constant $J$ is determined by the ratio \cite{kugel1982jahn}: $J\approx T^2/U$, 
where  $U$ and $T$ are the Coulomb interaction and electron hopping terms:
\begin{eqnarray}\label{localized and itinerant}
&&U=\frac{e^2}{4\pi\varepsilon_0l_0}\int\frac{\vert \psi_{n,\sigma}(\textbf{r}_1)\vert^2\, 
\vert \psi_{D,\sigma}(\textbf{r}_2)\vert^2}
{\vert\textbf{r}_1-\textbf{r}_2\vert+\delta}\, d\textbf{r}_1\, d\textbf{r}_2,
\nonumber \\
&& T=E\int\psi_{n,\sigma}^\dag (\textbf{r})\, \psi_{D,\sigma}(\textbf{r})\, d\textbf{r}.
\end{eqnarray}
Here $\hat\bsigma _B$ refers to the Pauli vector of the spin of quantum dot and $\delta $ is the cut-off of electron-electron interaction, $\hat\bsigma _A$ is the Pauli vector of the spin of itinerant electron in altermagnet, $R$ is the radius of the quantum dot. In what follows we set dimensionless Hamiltonian via $\hat H\equiv\hat H/J$, $J=1$. \vspace{0.2cm}\\
The initial wave function of the bipartite system is a product of two wave functions $\psi_{n,\sigma}(\textbf{r}_A)$ and $\psi_{D,\sigma}(\textbf{r}_B)=\psi(\textbf{r}_B)\vert 0\rangle_B$. In what follows, for brevity, we use the notations $\textbf{r}_A\equiv\textbf{r}$ and  $\textbf{r}_B\equiv\textbf{r}'$.
The scattering process involves two states of localized electron $\psi_{D,0}(\textbf{r})=\psi_D(\textbf{r})\vert 0\rangle$ (spin-up, $|0\rangle\equiv\ket{\uparrow}$ and  $\psi_{D,1}(\textbf{r})=\psi_D(\textbf{r})\vert 1\rangle$, spin-down, $|1\rangle\equiv\ket{\downarrow}$). The wave function of the two-electron system after scattering can be presented in the following general form:
\begin{equation}\label{two-electron system}
\Psi_{\sigma_1\sigma_2}(\textbf{r}_1,\textbf{r}_2)=\psi_{0}^{(+)}(\textbf{r}_1)\psi_{D,0}(\textbf{r}_2)+\psi_{1}^{(+)}(\textbf{r}_1)\psi_{D,1}(\textbf{r}_2).
\end{equation}
Here $\psi_{n,0}^{(+)}$ and $\psi_{n,1}^{(+)}$ are the two-component spinors
\begin{equation}\label{two-component spinors}
\psi_{0}^{(+)}(\textbf{r})= \begin{pmatrix} \phi_0(\textbf{r}) \\  \chi_0(\textbf{r}) \end{pmatrix}, \qquad 
\psi_{1}^{(+)}(\textbf{r})= \begin{pmatrix} \phi_1(\textbf{r}) \\  \chi_1(\textbf{r})
\end{pmatrix}.
\end{equation}
In the momentum-space representation, 
$$
\psi_{0(1)}^{(+)}(\textbf{p})=\int d^2r e^{-i{\bf pr}}\psi_{0(1)}^{(+)}(\textbf{r}),
$$ 
spinors in the above equation are the solution of the following system of coupled  Lippmann–Schwinger integral equations ($n={\bf k}, \sigma=\uparrow$):
\begin{eqnarray}\label{coupled integral equations second1}
&&\begin{pmatrix} \phi_0(\textbf{p}) \\ \chi_0(\textbf{p}) \end{pmatrix} =(2\pi)^2\delta({\bf p}-{\bf k})\left(\alpha\begin{pmatrix} 1 \\ 0 \end{pmatrix} +\beta\begin{pmatrix} 0 \\ 1 \end{pmatrix}\right)+\nonumber\\
&&\alpha\int \frac{d^2q}{(2\pi)^2}\,\hat{G}^{(+)}(\textbf{p};E)
\hat{V}_{00}(q)\begin{pmatrix} \phi_0(\textbf{p}-\textbf{q}) \\  \chi_0(\textbf{p}-\textbf{q}) \end{pmatrix}\nonumber\\
&&+\beta\int \frac{d^2q}{(2\pi)^2}\,\hat{G}^{(+)}(\textbf{p};E)
\hat{V}_{01}(\textbf{q})\begin{pmatrix} \phi_1(\textbf{p}-\textbf{q}) \\  \chi_1(\textbf{p}-\textbf{q}) \end{pmatrix},
\end{eqnarray}
and
\begin{eqnarray}\label{coupled integral equations second2}
&&\begin{pmatrix} \phi_1(\textbf{p}) \\ \chi_1(\textbf{p}) \end{pmatrix} =\nonumber\\
&&\alpha\int \frac{d^2q}{(2\pi)^2}\,\hat{G}^{(+)}(\textbf{p};E-2B)\hat{V}_{10}(\textbf{q})\begin{pmatrix} \phi_0(\textbf{p}-\textbf{q}) \\  \chi_0(\textbf{p}-\textbf{q}) \end{pmatrix}+\nonumber\\
&&\beta\int \frac{d^2q}{(2\pi)^2}\,\hat{G}^{(+)}(\textbf{p};E-2B)
\hat{V}_{11}(\textbf{r}')\begin{pmatrix} \phi_1(\textbf{p}-\textbf{q}) \\  \chi_1(\textbf{p}-\textbf{q}) \end{pmatrix}.
\end{eqnarray}
The Green function has the form: 
\begin{eqnarray}\label{The Green function}
&&\hat{G}^{(+)}(\textbf{p};E)=\frac{1}{2}\,\frac{\hat{I}+\hat{\sigma_z}}{E-E_{{\bf k},\uparrow}+i0}+\nonumber\\
&&\frac{1}{2}\,\frac{\hat{I}-\hat{\sigma_z}}{E-E_{{\bf k},\downarrow}+i0},
\end{eqnarray}
where $E_{k,\uparrow, \downarrow}=\varepsilon_k\pm\alpha_e k_xk_y$ and matrix elements read: $\hat{V}_{00}(\textbf{q})=Je^{-q^2l_B^2/2}\hat\sigma_z$, $\hat{V}_{01}(\textbf{q})=Je^{-q^2l_B^2/2}(\hat\sigma_x-i\hat\sigma_y)$, $\hat{V}_{10}(\textbf{q})=Je^{-q^2l_B^2/2}(\hat\sigma_x+i\hat\sigma_y)$, $\hat{V}_{11}(\textbf{r}_A)=-Je^{-q^2l_B^2/2}\hat\sigma_z$.\vspace{0.3cm}\\

\subsection{Calculation of integrals}

Fourier transformation of the wave functions (3) has the form
\begin{eqnarray}
&&\psi _0({\bf r})
=\int \frac{d^2p}{(2\pi )^2}\, e^{i{\bf p}\cdot {\bf r}}\,
\psi _0({\bf p})
=\alpha e^{i{\bf k}\cdot {\bf r}}
\begin{pmatrix} 1 \\ 0 \end{pmatrix}
+\beta e^{i{\bf k}\cdot {\bf r}}
\begin{pmatrix} 0 \\ 1 \end{pmatrix}\nonumber\\
&&+\alpha J\int \frac{d^2p}{(2\pi )^2}\, 
e^{i{\bf p}\cdot {\bf r}}\,
\frac{e^{-({\bf p-k})^2l_B^2/2}}
{E-E_{{\bf p},\uparrow}+i0}\,
\begin{pmatrix} 1 \\ 0 \end{pmatrix},
\end{eqnarray} 
\begin{eqnarray}
&&\psi _1({\bf r})
=\int \frac{d^2p}{(2\pi )^2}\, e^{i{\bf p}\cdot {\bf r}}\,
\psi _1({\bf p})
=\nonumber\\
&&\beta J\int \frac{d^2p}{(2\pi )^2}\, 
e^{i{\bf p}\cdot {\bf r}}\,
\frac{e^{-({\bf p-k})^2l_B^2/2}}
{E-2B-E_{{\bf p},\downarrow}+i0}\,
\begin{pmatrix} 0 \\ 1 \end{pmatrix}.
\end{eqnarray} 
As we see, we have to calculate two integrals
\begin{eqnarray}
\label{8}
&&I_1({\bf r})
=\int \frac{d^2p}{(2\pi )^2}\, 
e^{i{\bf p}\cdot {\bf r}}\,
\frac{e^{-a({\bf p-k})^2}}
{E-E_{{\bf p},\uparrow}+i0},
\\
&&I_2({\bf r})
=\int \frac{d^2p}{(2\pi )^2}\, 
e^{i{\bf p}\cdot {\bf r}}\,
\frac{e^{-a({\bf p-k})^2}}
{E-2B-E_{{\bf p},\downarrow}+i0}.
\end{eqnarray}
where $a=l_B^2/2$ and $E_{\bf p \uparrow, \downarrow }=\frac{\hbar ^2p^2}{2m_e}\pm \alpha _e p_xp_y$.\vspace{0.2cm}\\
Let us use an integral representation of the function:
\begin{eqnarray}
\label{10}
e^{-a({\bf p-k})^2}=\frac1{2\pi a}\int d^2r'\, e^{-i({\bf p-k})\cdot {\bf r'}}\, e^{-r'^2/2a}.
\end{eqnarray}
Then we get
\begin{eqnarray}
\label{11}
&&I_1({\bf r})
=\frac1{2\pi a}\int d^2r'\, e^{i{\bf k}\cdot {\bf r'}}\, e^{-r'^2/2a}\times\nonumber\\ 
&&\int \frac{d^2p}{(2\pi )^2}\, 
\frac{e^{i{\bf p}\cdot ({\bf r}-{\bf r'})}}
{E-E_{{\bf p},\uparrow}+i0}.
\end{eqnarray}
The integral over $\bf p$ in Eq.(\ref{11}) is
\begin{widetext}
\begin{eqnarray}
\label{12}
&&I_3({\bf r-r'})
=\int \frac{d^2p}{(2\pi )^2}\, 
\frac{e^{i{\bf p}\cdot ({\bf r-r'})}}
{E-E_{{\bf p},\uparrow}+i0}
=\frac1{4\pi ^2}\int dp_x dp_y\, 
\frac{e^{ip_x(r_x-r'_x)+ip_y(r_y-r'_y)}}
{E-\hbar ^2p^2/2m_e-\alpha _ep_xp_y+i0}=
\\
&&-\frac{m_e}{2\pi ^2\hbar ^2}
\int _{-\infty }^\infty dp_y\, e^{ip_y(r_y-r'_y)}
\int _{-\infty }^\infty dp_x\,
\frac{e^{ip_x(r_x-r'_x)}}
{(p_x-p_{x1})(p_x-p_{x2})},\nonumber
\end{eqnarray} 
\end{widetext}
where
\begin{eqnarray}
\label{14}
p_{x1,2}=-\xi p_y
\pm \Big( \xi ^2p_y^2-p_y^2
+\frac{2m_eE}{\hbar ^2}+i0\Big) ^{1/2}.
\end{eqnarray}
and we introduced notation $\xi = \alpha _e/\alpha_c$, where $\alpha_c = \hbar ^2/m_e$ is the maximum value of $\alpha_e$ \cite{PhysRevB.108.L140408}.
Calculating the integral over $p_x$ along the contour in complex $p_x$-plane, that can be closed in upper halfplane for $r_x>r'_x$ and in the lower halfplane for $r_x<r'_x$ we obtain
\begin{eqnarray}
\label{15}
&&\int _{-\infty }^\infty dp_x\,
\frac{e^{ip_x(r_x-r'_x)}}
{(p_x-p_{x1})(p_x-p_{x2})}
=\nonumber\\
&&\frac{2\pi i}{p_{x1}-p_{x2}}\left\{ 
\begin{array}{cc} e^{ip_{x1}(r_x-r'_x)}, & r_x>r'_x \\
e^{ip_{x2}(r_x-r'_x)}, & r_x<r'_x	
\end{array} \right. .
\end{eqnarray}
Note that this solution will be substituted to Eq.(\ref{11}). This means that we can take in Eq.(\ref{12}) $r_x>r'_x$ as it  corresponds to ${\bf r}$  outside the dot.
Substituting (15) with $r_x>r'_x$ to (13) we obtain
\begin{eqnarray}
\label{16}
&&I_3({\bf r-r'})
=\nonumber\\
&&-\frac{im_e}{2\pi \hbar ^2}
\int _{-\infty }^\infty dp_y\, 
\frac{e^{ip_y(r_y-r'_y)}\, e^{ip_{x1}(r_x-r'_x)}}
{\Big( \xi ^2p_y^2-p_y^2
+\frac{2m_eE}{\hbar ^2}+i0\Big) ^{1/2}}.
\end{eqnarray}  
Now we can substitute Eq.(\ref{16}) into Eq.(\ref{11})
\begin{widetext}
\begin{eqnarray}
\label{18}
&&I_1({\bf r})
=-\frac{im_e}{4\pi ^2\hbar ^2a}
\int _{-\infty }^\infty dp_y\,
\frac{e^{ip_yr_y}\, e^{ip_{x1}r_x}}
{\Big( \xi ^2p_y^2-p_y^2
+\frac{2m_eE}{\hbar ^2}+i0\Big) ^{1/2}}
\int d^2r'\, e^{-r'^2/2a} \,
e^{-i[(p_{x1}-k_x)\, r'_x+(p_y-k_y)\, r'_y]}
=\nonumber\\
&&-\frac{im_e}{2\pi \hbar ^2}
\int _{-\infty }^\infty dp_y\;
e^{ip_{x1}r_x+ip_yr_y}\; 
\frac{e^{-a[(p_{x1}-k_x)^2+(p_y-k_y)^2]}}
{\Big( \xi ^2p_y^2-p_y^2
+\frac{2m_eE}{\hbar ^2}+i0\Big) ^{1/2}}.
\end{eqnarray}
\end{widetext}
If we drop $i0$ in the denominator of (13), the function under integral is divergent, but the integral convergent. Thus, we can drop $i0$. Then we get
\begin{eqnarray}
\label{19}
&&I_1({\bf r})
=-\frac{im_e}{2\pi \hbar ^2\sqrt{1-\xi ^2}}\times\nonumber\\
&&\int _{-\infty }^\infty dp_y\;
e^{ip_{x1}r_x+ip_yr_y}\; 
\frac{e^{-a[(p_{x1}-k_x)^2+(p_y-k_y)^2]}}
{\sqrt{(\kappa -p_y)(\kappa +p_y)}},
\end{eqnarray}
where (we assume $E>0$ and $\xi <1$) 
\begin{eqnarray}
\label{20}
\kappa ^2=\frac{2m_eE}{\hbar ^2(1-\xi ^2)}.	
\end{eqnarray}
Integral Eq.(\ref{19}) can be evaluated approximately. Since we are interested in large $r\gg l_B$ (far from the dot), the main contribution in Eq.(\ref{19}) is from rather small $|p_y|<|r_y|^{-1}$. We assume $\kappa r\gg 1$. In this case, $(\kappa ^2-p_y^2)^{1/2}\simeq \kappa \; $ in the denominator of Eq.(\ref{19}). Besides, in the limit of $\kappa r\gg 1$, we also have $p_{x1,2}\simeq -\xi p_y\pm \kappa _0$, where $\kappa _0=\sqrt{2m_eE}/\hbar $. Hence, from Eq.(\ref{19}) we get
\begin{widetext}
\begin{eqnarray}
\label{21}
&&I_1({\bf r})
\simeq -\frac{im_ee^{i\kappa _0 r_x}}{2\pi \hbar ^2\kappa _0}
\int _{-\infty }^\infty dp_y\;
e^{-i\xi p_yr_x+ip_yr_y}\; 
e^{-a[(-\xi   p_y+\kappa _0-k_x)^2+(p_y-k_y)^2]}
\nonumber\\
&&=-\frac{im_ee^{i[\kappa _0 r_x -\frac{\xi r_x-r_y}{1+\xi^2}(k_y+\xi(\kappa_0-k_x))]}}
{2\sqrt{\pi a(\xi ^2+1)}\, \hbar ^2\kappa _0}\;
\exp\Big\{- \frac{(\xi r_x-r_y)^2+4a^2(\xi k_y-(\kappa_0-k_x))^2}{4a(\xi ^2+1)}\Big\}.
\end{eqnarray}
\end{widetext}
Integral $I_2({\bf r})$ differs from $I_1({\bf r})$ by the sign of $\alpha _e$. Correspondingly, we need to change the sign of $\xi $ in (23) to get the expression for $I_2({\bf r})$. Besides, we have to change $E\to E-2B$. Thus, we obtain 
\begin{eqnarray}
\label{24}
&&I_2({\bf r})
\simeq -\frac{im_ee^{i[\kappa _1 r_x +\frac{\xi r_x+r_y}{1+\xi^2}(k_y-\xi(\kappa_1-k_x))]}}
{2\sqrt{\pi a(\xi ^2+1)}\, \hbar ^2\kappa _1}\nonumber\\
&&\exp\Big\{- \frac{(\xi r_x+r_y)^2+4a^2(\xi k_y+(\kappa_1-k_x))^2}{4a(\xi ^2+1)}\Big\}.
\end{eqnarray}
where $\kappa _1=\sqrt{2m_e(E-2B)}/\hbar $.  
The obtained result is valid for  $\kappa r\gg 1$, $\kappa _1r\gg 1$ and magnetic fields $B<E/2$.

\section{Post-scattering density matrix}

Taking into account the initial density matrix 
\begin{eqnarray}\label{the initial density matrix}
\hat\rho_0=\left(\psi_{T,\sigma}(\textbf{r})\psi_{D}\right)
\left(\psi_{T,\sigma}(\textbf{r})\psi_{D}\right)^\dagger\otimes\ket{0}\bra{0}_B,
\end{eqnarray}
and the wave function of the system after scattering Eq.(\ref{two-electron system}), we define two random unitary matrices through following formulae 
\begin{eqnarray}\label{through following formulae first}
&&\hat\rho=\ket{\Psi_{\sigma_1\sigma_2}(\mathbf{r},\mathbf{r}')}\bra{\Psi_{\sigma_1\sigma_2}(\mathbf{r},\mathbf{r}')}=\nonumber\\
&&\frac{1}{Tr\left(\hat U\hat\rho_0\hat U^\dagger\right)}\hat U\hat\rho_0\hat U^\dagger,
\end{eqnarray}
\begin{eqnarray}\label{through following formulae second}
&&\hat\rho_A=\frac{1}{Tr\left(\hat U_A\hat\rho_0\hat U_A^\dagger\right)}\hat U_A\hat\rho_0\hat U_A^\dagger.
\end{eqnarray}
Here index $\textbf{A}$ means that gate transformation $\hat U_A$ acts only on the qubit $\textbf{A}$ while the qubit of the localized electron $\textbf{B}$ is frozen by applied strong magnetic field. The explicit form of the density matrices read:
\begin{eqnarray}\label{explicit form of the density matrices first}
&&\hat\rho=\rho_{11}\ket{0}\bra{0}_A\otimes\ket{0}\bra{0}_B+\rho_{22}\ket{1}\bra{1}_A\otimes\ket{0}\bra{0}_B+\nonumber\\
&&\rho_{44}\ket{1}\bra{1}_A\otimes\ket{1}\bra{1}_B+\rho_{12}\ket{0}\bra{1}_A\otimes\ket{0}\bra{0}_B+\nonumber\\
&&\rho_{14}\ket{0}\bra{1}_A\otimes\ket{0}\bra{1}_B+\rho_{21}\ket{1}\bra{0}_A\otimes\ket{0}\bra{0}_B+\nonumber\\
&&\rho_{41}\ket{1}\bra{0}_A\otimes\ket{1}\bra{0}_B+\rho_{24}\ket{1}\bra{1}_A\otimes\ket{0}\bra{1}_B+\nonumber\\
&&\rho_{42}\ket{1}\bra{1}_A\otimes\ket{1}\bra{0}_B,
\end{eqnarray}
where
\begin{eqnarray}\label{explicit form of the density matrices second}
&&\rho_{11}=\frac{\alpha^2|e^{i\vec{k}\vec{r}}+JI_1(\textbf{r})|^2}{\alpha^2|e^{i\vec{k}\vec{r}}+JI_1(\textbf{r})|^2+\beta^2+\beta^2J^2|I_2(\textbf{r})|^2},\nonumber\\
&&\rho_{22}=\frac{\beta^2}{\alpha^2|e^{i\vec{k}\vec{r}}+JI_1(\textbf{r})|^2+\beta^2+\beta^2J^2|I_2(\textbf{r})|^2},\nonumber\\
&&\rho_{44}=\frac{\beta^2J^2|I_2(\textbf{r})|^2}{\alpha^2|e^{i\vec{k}\vec{r}}+JI_1(\textbf{r})|^2+\beta^2+\beta^2J^2|I_2(\textbf{r})|^2},
\end{eqnarray}
and
\begin{eqnarray}\label{second explicit form of the density matrices second}
&&\rho_{12}=\frac{\alpha\beta(1+Je^{-i\vec{k}\vec{r}}I_1(\textbf{r}))}{\alpha^2|e^{i\vec{k}\vec{r}}+JI_1(\textbf{r})|^2+\beta^2+\beta^2J^2|I_2(\textbf{r})|^2},\nonumber\\
&&\rho_{14}=\frac{\alpha\beta JI^*_2(\textbf{r})(e^{i\vec{k}\vec{r}}+JI_1(\textbf{r}))}{\alpha^2|e^{i\vec{k}\vec{r}}+JI_1(\textbf{r})|^2+\beta^2+\beta^2J^2|I_2(\textbf{r})|^2},\nonumber\\
&&\rho_{24}=\frac{\beta^2Je^{i\vec{k}\vec{r}}I_2^*(\textbf{r})}{\alpha^2|e^{i\vec{k}\vec{r}}+JI_1(\textbf{r})|^2+\beta^2+\beta^2J^2|I_2(\textbf{r})|^2},
\end{eqnarray}
$\rho_{12}=\rho^*_{21}$,  $\rho_{14}=\rho^*_{41}$,  $\rho_{24}=\rho^*_{42}$.  For the second matrix we deduce:
\begin{eqnarray}\label{explicit form of the density matrices third}
&&\hat\varrho_A=\rho_{A11}\ket{0}\bra{0}_A\otimes\ket{0}\bra{0}_B+\rho_{A22}\ket{1}\bra{1}_A\otimes\ket{0}\bra{0}_B+\nonumber\\
&&\rho_{A12}\ket{0}\bra{1}_A\otimes\ket{0}\bra{0}_B+\rho_{A21}\ket{1}\bra{0}_A\otimes\ket{0}\bra{0}_B,
\end{eqnarray}
and
\begin{eqnarray}\label{explicit form of the density matrices second}
&&\rho_{A11}=\frac{\alpha^2|e^{i\vec{k}\vec{r}}+JI_1(\textbf{r})|^2}{\alpha^2|e^{i\vec{k}\vec{r}}+JI_1(\textbf{r})|^2+\beta^2},\nonumber\\
&&\rho_{A22}=\frac{\beta^2}{\alpha^2|e^{i\vec{k}\vec{r}}+JI_1(\textbf{r})|^2+\beta^2},\nonumber\\
&&\rho_{A12}=\frac{\alpha\beta(1+Je^{-i\vec{k}\vec{r}}I_1(\textbf{r}))}{\alpha^2|e^{i\vec{k}\vec{r}}+JI_1(\textbf{r})|^2+\beta^2},\nonumber\\
\end{eqnarray}
and $\rho_{A12}=\rho^*_{A21}$.\vspace{0.3cm}\\

\section{Concurrence and Fidelity}
\label{sec:Concurrence}

We exploit definition of concurrence \cite{PhysRevLett.80.2245}
$\mathcal{C}=\vert\bra{\psi}\hat\sigma_y\otimes\hat\sigma_Y\ket{\psi^*}\vert$, where $\hat\sigma_y\otimes\hat\sigma_Y$ is the direct product of Pauli matrices and calculate entanglement between two electrons after scattering:
\begin{eqnarray}\label{entanglement between two electrons}
\mathcal{C}=\frac{2\alpha\beta J\vert \left(e^{-i\vec{k}\vec{r}}+JI^*_1(\textbf{r})\right) I_2^*(\textbf{r})\vert. }{\alpha^2|e^{i\vec{k}\vec{r}}+JI_1(\textbf{r})|^2+\beta^2+\beta^2J^2|I_2(\textbf{r})|^2}.
\end{eqnarray}
In what follows we explore concurrence in absence of disorder. 
For the fidelity between Bell's state $\ket{\Phi}_{AB}^+$ and scattered state $\hat\rho$ we deduce:
\begin{eqnarray}\label{Fidelity}
&&\mathcal{F}\left(\hat\rho, \ket{\Phi}_{AB}^+ \right)=\frac{1}{2}\left(\rho_{11}+\rho_{44}+\rho_{14}+\rho_{41}\right).
\end{eqnarray}
When $\alpha=\beta$, $J=1$,  Eq.(\ref{Fidelity}) takes the form:
\begin{eqnarray}\label{2Fidelity}
&&\mathcal{F}_{\alpha=\beta}\left(\hat\rho, \ket{\Phi}_{AB}^+ \right)=\frac{|I_4(\textbf{r})|^2+|I_2(\textbf{r})|^2}{2\left(|I_4(\textbf{r})|^2+|I_2(\textbf{r})|^2+1\right)}+\nonumber\\
&&\frac{\text{Re}(I_2^*(\textbf{r})I_4(\textbf{r}))}{\left(|I_4(\textbf{r})|^2+|I_2(\textbf{r})|^2+1\right)},
\end{eqnarray}
where we introduced the notation $I_4(\textbf{\textbf{r}})=e^{i\textbf{k}\textbf{r}}+I_1(\textbf{r})$. When $\alpha+1$, $\beta=0$ fidelity takes the value $\mathcal{F}\left(\hat\rho, \ket{\Phi}_{AB}^+ \right)=1/2$.
\begin{figure*} 
\includegraphics[width=\textwidth]{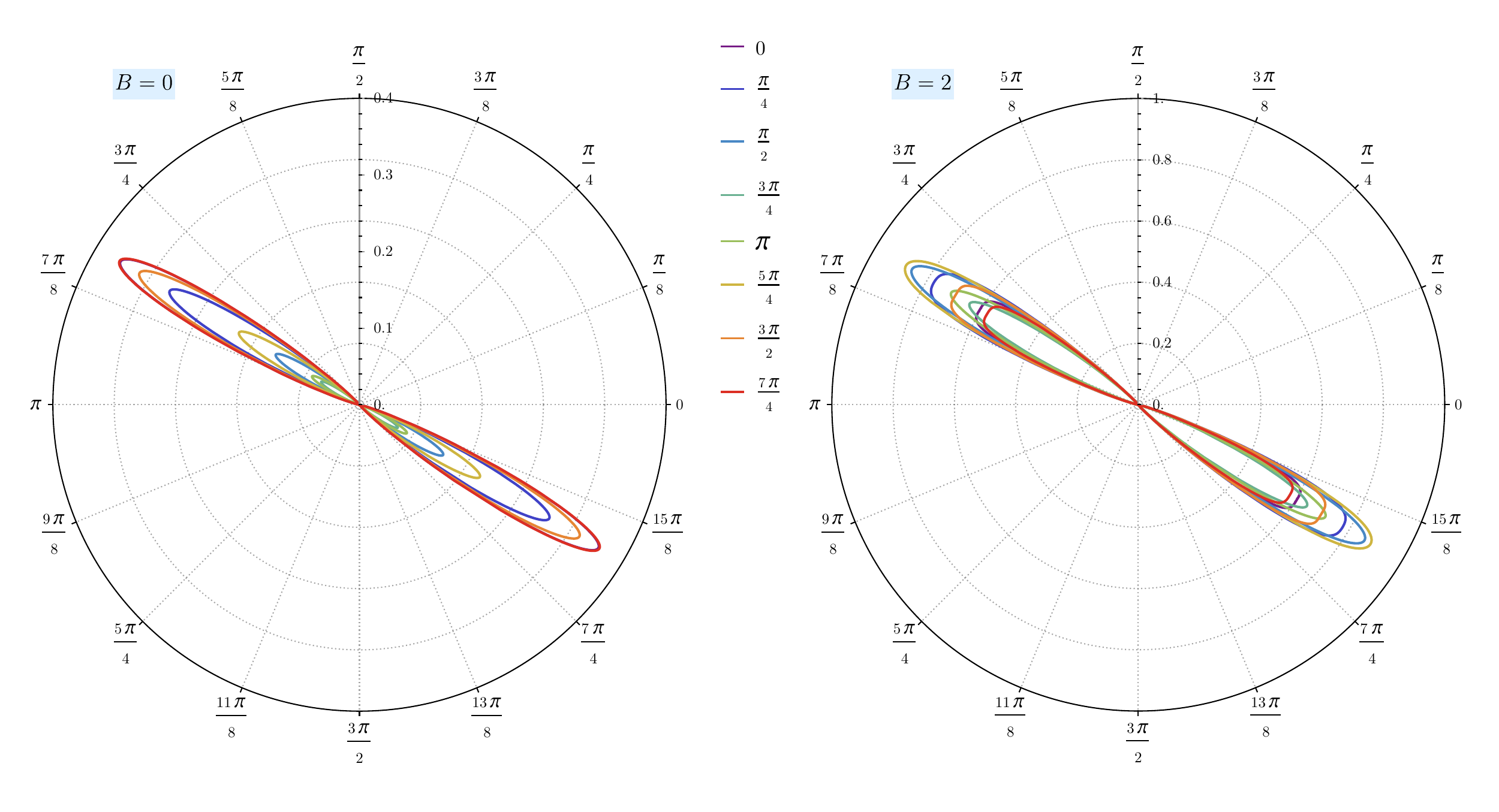}
\caption{The spatially anisotropic distribution of concurrence $\mathcal{C}(\varphi)$, in the polar coordinate system $r_x=r\cos\varphi$, $r_y=r\sin\varphi$,  for equally distributed  spin up/down incident states $\alpha^2=\beta^2=1/2$ of the itinerant electron. We present results for different $\theta$: $k_x=k\cos\theta$, $k_y=k\sin\theta$, for zero $B=0$ (the left plot) and non-zero $B=2$T magnetic fields (the right plot). The values of parameters read $m=m_e$, $\alpha_e/\alpha_c=0.6$, $k\equiv l_0k=1$, $E=0.5$ meV, $r\equiv r/l_0=10$ and $l_0=(\hbar/m_e\omega_0)^{1/2}=10$nm is the confinement length, $\alpha_c=\hbar^2/m_e$. Different colors annotate values of $\theta$.} 
\label{fig:down}
\end{figure*}
\begin{figure*} 
\includegraphics[width=\textwidth]{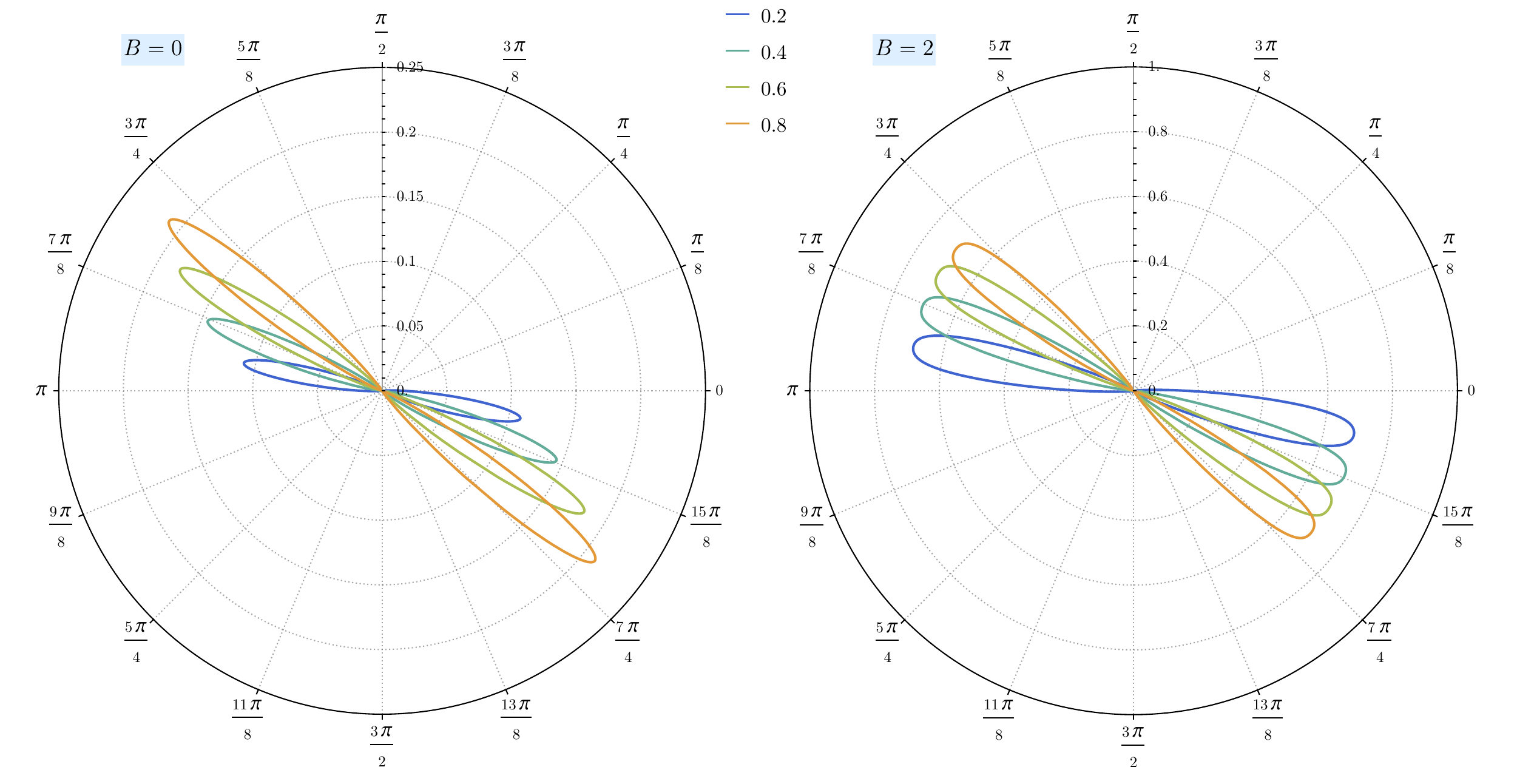}
\caption{The rotation of concurrence $\mathcal{C}(\varphi)$ in the polar coordinate system $r_x=r\cos\varphi$, $r_y=r\sin\varphi$, for different values of the spin-orbit constant constant $\alpha_e$, $\alpha_e/\alpha_c=0.2;\,0.4;\,0.6;\,0.8$ and equally distributed  spin up/down incident states $\alpha^2=\beta^2=1/2$ of the itinerant electron. The value of $\theta$: $k_x=k\cos\theta$, $k_y=k\sin\theta$, $\theta=5\pi/4$ . The following parameters are considered: $m=m_e$, $k\equiv l_0k=1$, $E=0.5$ meV, $r\equiv r/l_0=10$, $\alpha_c=\hbar^2/m_e$, the confinement length $l_0=10$ nm.  The magnetic fields $B=0$ (the left plot) and $B=2$T (the right plot). Different colors annotate values of $\alpha_e/\alpha_c$.} 
\label{fig:equal}
\end{figure*}
\begin{figure*}
\includegraphics[width=\textwidth]{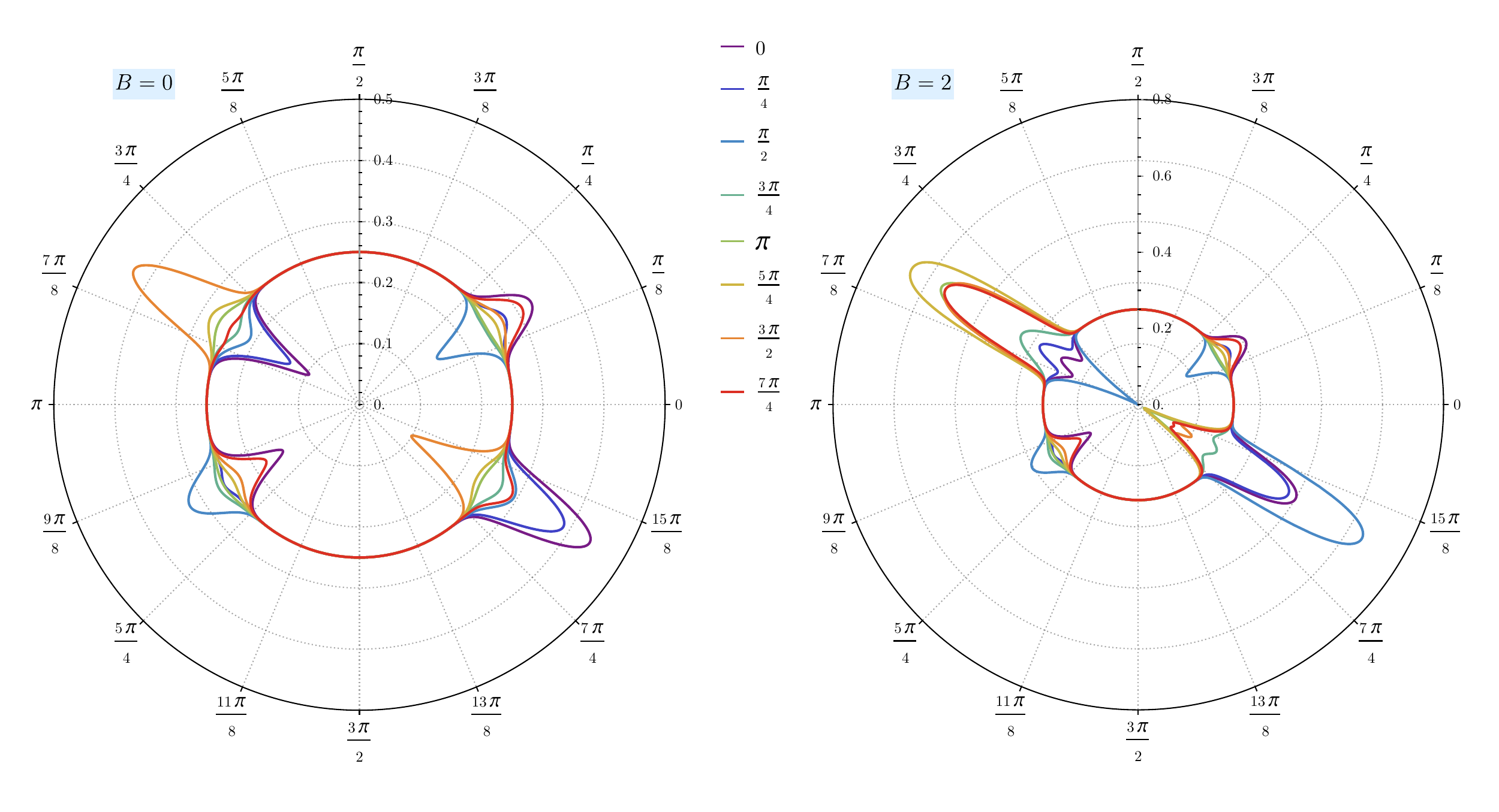}
\caption{The spatially anisotropic distribution of fidelity $\mathcal{F}(\varphi)$ in the polar coordinate system $r_x=r\cos\varphi$, $r_y=r\sin\varphi$, for equally distributed  spin up/down incident states $\alpha^2=\beta^2=1/2$ of the itinerant electron. We present results for different $\theta$: $k_x=k\cos\theta$, $k_y=k\sin\theta$, zero (the left plot) and non-zero magnetic fields $B=2$T (the right plot). The values of parameters $m=m_e$, $\alpha_e/\alpha_c=0.6$, $\alpha_c=\hbar^2/m_e$, $k\equiv l_0k=1$, $E=0.5$ meV, $r\equiv r/l_0=10$. The confinement length $l_0=10$nm. Different colors annotate values of $\theta$.} 
\label{fig:up}
\end{figure*}
In Fig.\ref{fig:down}, we plot the spatially anisotropic distribution of entanglement $r_x=r\cos\varphi$, $r_y=r\sin\varphi$,  for equally distributed  spin up/down incident states $\alpha^2=\beta^2=1/2$ of the itinerant electron beam. We present results for different $\theta$: $k_x=k\cos\theta$, $k_y=k\sin\theta$, for zero and non-zero magnetic fields and other parameters as they are shown in the figure caption. As we see from Fig.\ref{fig:down}
magnetic field $B$ substantially enhances the value of concurrence, from $\mathcal{C}(\varphi)=0.35$ up to the $\mathcal{C}(\varphi)=0.9$. Besides, due to the strong SOI in the system, concurrence strongly depends on the momentum $\textbf{k}$ of the itinerant electron. We see an interesting physical effect in Fig.\ref{fig:equal}. Spatially anisotropic distribution of concurrence depends on the values of altermagnetic SOI constant $\alpha_e$. In the case of the weak SOI, constant concurrence rotates towards small scattering angles, meaning $\varphi=0,\pi$, $r_x\approx \pm r$,  $r_y\approx \pm 0$. 
In Fig.\ref{fig:up}, we plot the spatially resolved anisotropic distribution of Fidelity between the post-scattering state and Bell's state $\ket{\Phi}^+_{AB}, $ which is our desired target state. As we can see, akin to the concurrence, Fidelity's maximal value rotates upon the change in the spin-orbit constant. Besides, the value of Fidelity reaches $70\%$, which is a quite reasonable value for the gate $\left\lbrace \text{Hadamard}_A\circ\text{CNOT}_B\right\rbrace $.

\section{Rényi entropy}
\label{sec:Rényi entropy}

\begin{figure*}
\includegraphics[width=\textwidth]{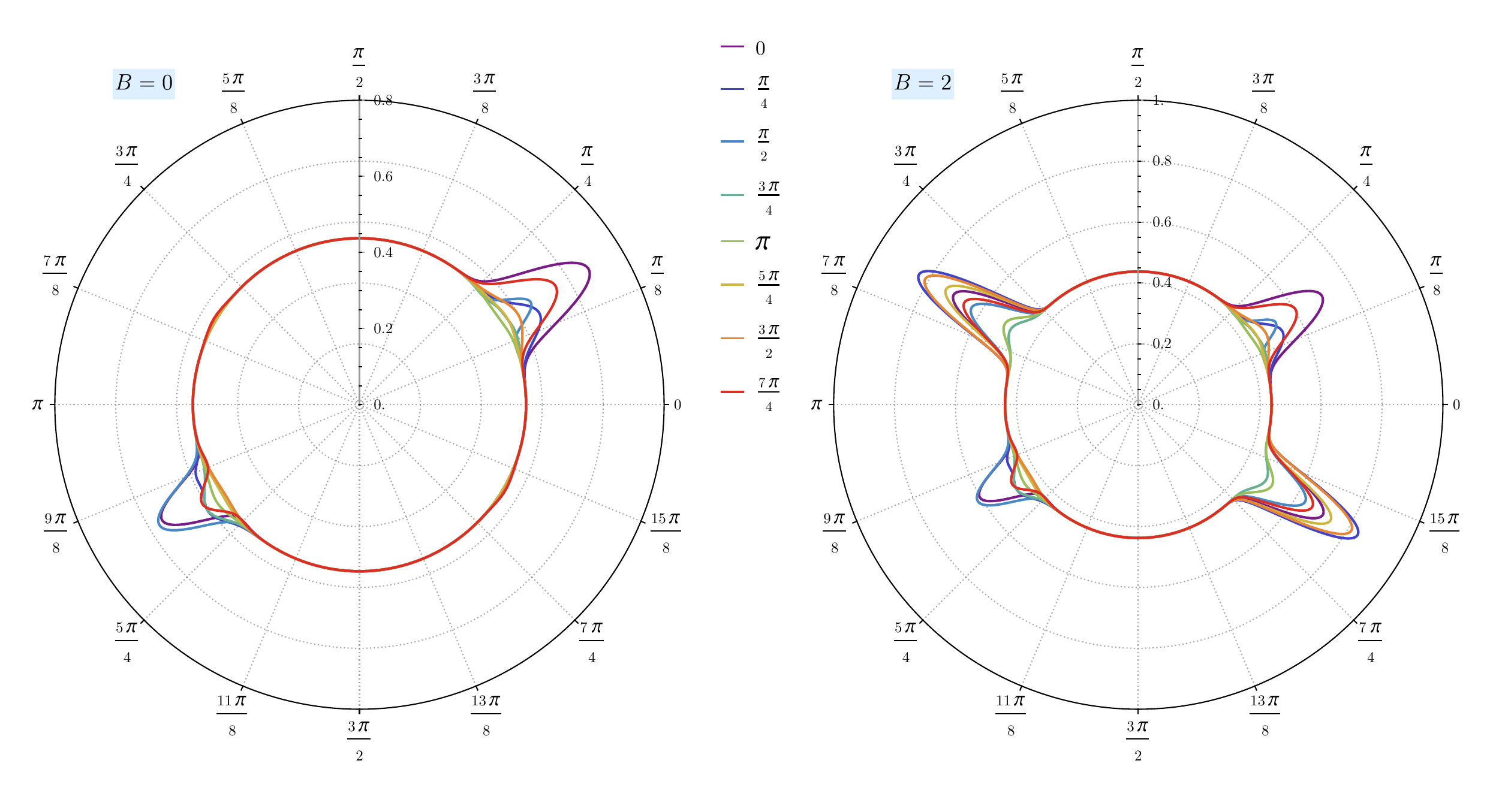}
\caption{The spatially anisotropic Renyi entropy $S_2\left(\hat\varrho[\varphi]\right)=-\log_2Tr\left[ \hat \varrho^2\right]$ in the polar coordinate system $r_x=r\cos\varphi$, $r_y=r\sin\varphi$, for equally distributed  spin up/down incident states $\alpha^2=\beta^2=1/2$ of the itinerant electron. The parameters of the distribution function $\mathcal{P}(\alpha_e)$: $\bar\xi=1$, $\Delta \xi=0.1$  We present results for different $\theta$: $k_x=k\cos\theta$, $k_y=k\sin\theta$, zero (the left plot) and non-zero magnetic fields $B=2$T (the right plot). The values of parameters $m=m_e$, $k\equiv l_0k=1$, $E=0.5$ meV, $r\equiv r/l_0=10$. The confinement length $l_0=10$nm. Different colors annotate values of $\theta$.} 
\label{fig:renyi}
\end{figure*}

We proceed and calculate Rényi entropy of the bipartite system after scattering taking into account disorder in the system and randomness of the spin-orbit interaction constant. 
Following \cite{PhysRevA.99.052323}, we present bipartite density matrix through the Pauli strings:
\begin{eqnarray}\label{Pauli strings}
&&\hat \varrho=\frac{1}{4}\sum\limits_{\mu,\nu=0}^3\mathcal{R}_{\mu\nu}D_{\mu\nu},\nonumber\\
&&\mathcal{R}_{\mu\nu}=Tr\left[\hat\varrho D_{\mu\nu}\right],~~\hat\sigma_0=\mathcal{I}_{A,B}.\nonumber\\
&&D_{\mu\nu}=\hat\sigma^\mu_A\otimes\hat\sigma^\nu_B, 
\end{eqnarray}
where $D_{\mu\nu}$ is the Dirac matrix \cite{PhysRevA.93.062320}. After laborious calculations we deduce:
\begin{widetext}
\begin{eqnarray}
&&\mathbf{\mathcal{R}}=\nonumber\\
&&\begin{bmatrix}
1 & 2\text{Re}(\rho_{12}) & -2\text{Im}(\rho_{12}) & \rho^2_{11}-\rho^2_{22}-\rho^2_{44}\\
2\text{Re}(\rho_{24}) &  2\text{Re}(\rho_{14}) & -2\text{Im}(\rho_{14}) & -2\text{Re}(\rho_{24})\\
-2\text{Im}(\rho_{24}) & -2\text{Im}(\rho_{14}) & -2\text{Re}(\rho_{14}) & 2\text{Im}(\rho_{24})\\
\rho_{11}^2+\rho_{22}-\rho_{44} & 2\text{Re}(\rho_{12}) & -2\text{Im}(\rho_{12}) & \rho_{11}^2+\rho_{22}+\rho_{44}
\end{bmatrix}.
\end{eqnarray}
\end{widetext}
The quantity of interest, the second Rényi entropy is given as:
\begin{eqnarray}\label{Renyi entropy}
&& S_2\left(\hat\varrho\right)=-\log_2Tr\left[ \hat \varrho^2\right].   
\end{eqnarray}
Trace of the density matrix can be calculated through the following formula \cite{PhysRevA.99.052323}:
\begin{eqnarray}\label{Thrace of the density matrix}
&& Tr\left[ \hat \varrho^2\right]=\frac{1}{4}\bigg[ 1+3\left\langle \left(Z_u^A\right)^2\right\rangle+\nonumber\\
&&3\left\langle \left(Z_u^B\right)^2\right\rangle+ 9\left\langle \left(Z_u^{AB}\right)^2\right\rangle\bigg],  
\end{eqnarray}
where 
\begin{eqnarray}\label{the following quantities}
&& Z_u^A=Tr\left[\hat U_A \hat\varrho \hat U_A^\dagger\hat\sigma_z^A\otimes \mathcal{I}_B\right],\nonumber\\
&& Z_u^B=Tr\left[\hat U_B \hat\varrho \hat U_B^\dagger\mathcal{I}_A\otimes\hat\sigma_z^B\right],\nonumber\\
&& Z_u^{AB}=Tr\left[\hat U \hat\varrho \hat U^\dagger\hat\sigma_z^A\otimes\hat\sigma_z^B\right], 
\end{eqnarray}
and ensemble average moments $\langle ...\rangle$ is done for the random SOI. Here $Z_u^{AB}$ corresponds to the process when both spins of electrons are flipped after scattering process, while $Z_u^A$ and $Z_u^B$ correspond to the processes when only one spin is flipped. 
After cumbersome calculations we deduce:
\begin{widetext}
\begin{eqnarray}\label{post-scattering density matrix}
&& \hat U \hat\varrho \hat U^\dagger\hat\sigma_z^A\otimes\hat\sigma_z^B=\bigg(\rho_{11}\ket{0}\bra{0}_A\otimes\ket{0}\bra{0}_B-\rho_{22}\ket{0}\bra{0}_A\otimes\ket{1}\bra{1}_B+\rho_{44}\ket{1}\bra{1}_A\otimes\ket{1}\bra{1}_B-\nonumber\\
&&\rho_{12}\ket{0}\bra{0}_A\otimes\ket{0}\bra{1}_B+\rho_{14}\ket{0}\bra{1}_A\otimes\ket{0}\bra{1}_B+\rho_{21}\ket{0}\bra{0}_A\otimes\ket{1}\bra{0}_B-\rho_{24}\ket{0}\bra{1}_A\otimes\ket{1}\bra{1}_B\nonumber\\
&&+\rho_{41}\ket{1}\bra{0}_A\otimes\ket{1}\bra{0}_B-\rho_{42}\ket{1}\bra{0}_A\otimes\ket{1}\bra{1}_B\bigg).
\end{eqnarray}
\end{widetext}
Taking into account Eq(\ref{post-scattering density matrix}) we deduce:
\begin{eqnarray}\label{here two}
&& Z_u^{AB}=Tr\left[\hat U \hat\varrho \hat U^\dagger\hat\sigma_z^A\otimes\hat\sigma_z^B\right]=\rho_{11}-\rho_{22}+\rho_{44}.
\end{eqnarray}
If the spin of the quantum dot is frozen, similarly we obtain:
\begin{eqnarray}\label{spin of the quantum dot is frozen}
&&Z_u^A=\hat U_A \hat\varrho \hat U_A^\dagger\hat\sigma_z^A\otimes \mathcal{I}_B=\rho_{A11}\ket{0}\bra{0}_A\otimes\ket{0}\bra{0}_B.
\end{eqnarray}

As a next step we need to define mean values of squares of quantities Eq.(\ref{the following quantities}). 
These averaged values over random $\alpha_e=\alpha_c\xi$ are defined as follows:
\begin{eqnarray} 
\label{we deduce the explicit expression}
&& \left\langle \left(Z_u^A\right)^2\right\rangle=\int d\xi\mathcal{P}(\xi)\left(Z_u^A(\xi)\right)^2,\nonumber\\
&& \left\langle \left(Z_u^B\right)^2\right\rangle=0,\nonumber\\
&& \left\langle \left(Z_u^{AB}\right)^2\right\rangle=\int d\xi\mathcal{P}(\xi)\left(Z_u^{AB}(\xi)\right)^2.\\\nonumber
\end{eqnarray}
Here $\mathcal{P}(\xi)=\frac{1}{\Delta \xi\sqrt{2\pi}}\exp\left[-\frac{(\xi-\bar{\xi}_e)^2}{2(\Delta \xi)^2}\right] $ is the distribution function of the random SO interaction constant $v$. The Renyi entropy as a function of magnetic field $B$ is plotted in 
Fig.\ref{fig:renyi}. As we see Renyi entropy depends on the angles $\theta,\,\varphi$ and as well as concurrence has anisotropic character. 
As we see the maximal values of the  Renyi entropy are rotated by magnetic field $B$ in the $(k_x,k_y)$ plane.

\section{Conclusions}
\label{sec:Conclusions}
In the present work, we studied entanglement in the system of an altermagnetic and a quantum dot. In particular, we considered the case when the initially itinerant electron and the electron localized in the quantum dot are disentangled and become entangled after the spin-resolved scattering process. We analytically solved coupled  Lippmann–Schwinger integral equations and obtained the system's post-scattering density matrix. We found that concurrence in the system is highly anisotropic, and the region of non-zero concurrence rotates depending on the values of spin-orbit constant. Realistic physical systems, in many cases, are characterized by certain disorders due to lattice dislocations and impurities. To describe the effects of the disorder, we considered the random spin-orbit constant and calculated the Renyi entropy averaged over the randomness in the spin-orbit constant. Akin to the concurrence, the Renyi entropy also depends on the spin-orbit constant. We showed that due to the unique properties of an altermagnetic system, tuning the applied external magnetic field allows tailoring of the desired entangled state. Thus, the scattering process, in essence, mimics the Hadamard-CNOT Gate transformation, converting the initial disentangled state into the entangled state of Bell's state with more than 70 percent fidelity.

\bibliography{TEXT_KK_VD.LC}

\end{document}